\documentstyle[prb,aps,psfig,multicol]{revtex}

\newcommand{\bfar}{\mbox{$B_{\rm far}$}}

\newcommand{\Bvec}{\mbox{${\bf B}$}}
\newcommand{\Evec}{\mbox{${\bf E}$}}
\newcommand{\jvec}{\mbox{${\bf j}_e$}}
\newcommand{\rvec}{\mbox{${\bf r}$}}
\newcommand{\rpvec}{\mbox{${\bf r}'$}}

\newcommand{\be}{\begin{equation}}
\newcommand{\ee}{\end{equation}}
\newcommand{\barr}{\begin{eqnarray}}
\newcommand{\earr}{\end{eqnarray}}
\newcommand{\breakeq}{\nonumber \\ &&}

\begin{document}
\draft
\title{Light emission from a scanning tunneling microscope:
       Fully retarded calculation}
\author{Peter Johansson\cite{Email}}
\address{Department of Theoretical Physics, University of Lund,
         S\"olvegatan 14 A, S--223\,62 Lund, Sweden}
\date{\today}
\maketitle
\begin{abstract}
The light emission rate from a scanning tunneling microscope (STM)
scanning a noble metal surface is calculated taking retardation effects
into account.
As in our previous, non-retarded theory 
[Johansson, Monreal, and Apell, Phys.\ Rev.\ B {\bf 42}, 9210 (1990)], 
the STM tip is modeled by a sphere, and the dielectric 
properties of tip and sample are described by experimentally measured
dielectric functions.
The calculations are based on exact diffraction theory through
the vector equivalent of the Kirchoff integral.
The present  results are qualitatively similar to those of the 
non-retarded calculations.
The light emission spectra have pronounced resonance peaks due to the 
formation of a tip-induced plasmon mode localized to the cavity between 
the tip and the sample.
At a quantitative level, the effects of retardation are rather small 
as long as the sample material is Au or Cu, and the tip consists of W or Ir.
However, for Ag samples, in which the resistive losses are smaller,
the inclusion of retardation effects in the calculation leads to larger
changes: the resonance energy decreases by 0.2--0.3 eV, 
and the resonance broadens.
These changes improve the agreement with experiment.
For a Ag sample and an Ir tip,
the quantum efficiency  is $\approx$ 10$^{-4}$ emitted photons in the 
visible frequency range per tunneling electron.
A study of the energy dissipation into the tip and sample shows 
that in total about 1 \% of the electrons undergo inelastic processes
while tunneling.
\end{abstract}



\begin{multicols}{2}

\section{Introduction}

The scanning tunneling microscope (STM) 
has over the last 15 years developed 
into a standard instrument in surface science. 
A relatively small, but very interesting part of this development
is concerned with light emission from STM's.
This phenomenon, in which the tunneling electron
transfers energy to a photon,\cite{Lambe} has been observed on 
noble-metal surfaces by Gimzewski, Berndt, and 
co-workers,\cite{Reihl,EPL,PRL91,Richard,Auresolv}
and later by several other groups.\cite{Smol,Sivel,Ito,Umeno,Dumas}
Light emission was also observed 
from semiconductors,\cite{Abra,Santos1,Montel1,Montel2} 
from surfaces covered by molecules,\cite{C60} 
and from magnetic surfaces.\cite{Alv}
Other experiments have studied closely related phenomena.
To mention a few examples, the  STM tip can be used as a local
detector of photoelectrons in photoemission,\cite{STMPE} 
and if the STM is illuminated by laser light, the rectification 
current can be large enough to generate STM topographs.\cite{Vol}
Moreover, the interaction between propagating surface plasmons 
and an STM tip has been investigated.\cite{Plasmexp}

\begin{figure}
\centerline{
\psfig{file=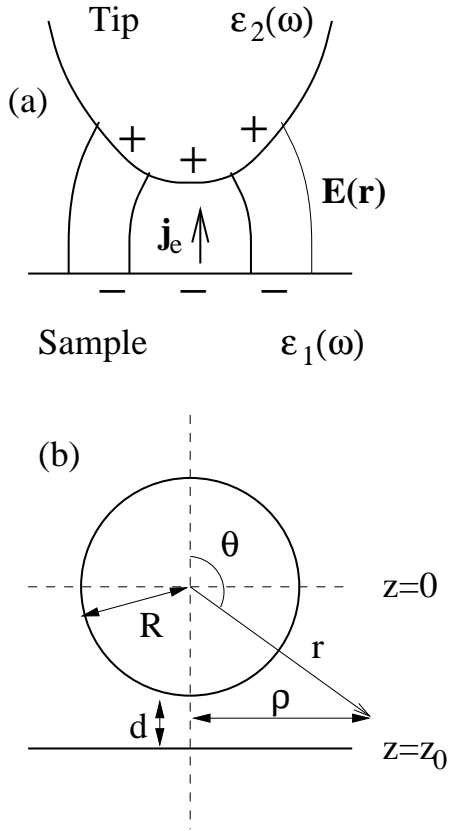,width=2.3in,%
bbllx=208 pt,bblly=211 pt,bburx=404 pt,bbury=580 pt}}
\vspace{0.1in}
\centerline{\caption{
Illustration of the tip-sample geometry (not to scale).
In (a) the current $\jvec$ exciting the system and the electric 
field associated with the tip-induced plasmon mode are shown.
In (b) the model geometry and 
coordinate system employed in the calculations are shown.
}
\label{Bild}
}
\end{figure}

Light emission from noble-metal surfaces 
has inspired a certain amount of
theoretical activity. \cite{JMA,ZPB,BNJP,Ueh,Denk,Madra,Downes} 
The most striking experimental results, the high 
(in this context) quantum 
efficiency of the process (up to 10$^{-3}$ photons per electron) and the 
characteristic resonances in the light emission spectra,
can be understood in the following 
way: \cite{JMA,ZPB,BNJP,Ueh,Denk,Madra,Downes,Rend,Malsh}
When the tip-sample separation is no more than 5--10 \AA, the surface 
plasmons on the two surfaces interact quite strongly and form an
interface-plasmon mode. 
This is illustrated schematically in Fig.\ \ref{Bild}.
The resonance in the light emission 
spectrum occurs at the frequency for which half a wavelength of the interface
mode fits into the ``cavity'' between tip and sample.
The charge oscillations associated with the interface plasmon have opposite
signs on the two electrodes. 
This has two major consequences:
(i) The charge on one surface gives a field that polarizes the second 
surface so that the polarizing field at the first surface increases, etc.
Thus, given the charge configuration of the interface mode, the enhancement
of the electromagnetic vacuum fluctuations is a quite natural consequence. 
(ii) Because the charges on the tip and the sample are opposite,
the interface plasmon is locally charge-neutral. 
This weakens the restoring forces compared with the case of a surface 
plasmon at an isolated interface, and the resonance peak is red-shifted
relative to the surface plasma frequency.
From this argument follows also a geometry-dependence of the resonance 
frequency. 
A larger tip radius  and a smaller 
tip-sample separation gives a more red-shifted resonance.

Model calculations based on the physical picture described above
gave results in rather good agreement with experiment,
both in terms of absolute intensities and spectral properties.
In these calculations, the complicated, extended geometrical shape 
of the STM tip was approximated by a sphere, and the electromagnetic field
near the model tip was calculated in the non-retarded limit.\cite{JMA,ZPB}
The justification for the approximations is that the region {\em between}
the tip and sample is most important for the effect. There the tip
shape is well approximated by a sphere, and all relevant distances are
small compared with the wavelength of the emitted photons.

However, in view of future developments, it is rather useful
to know how good these approximations are.
In this paper, we focus on the effects of retardation.
The calculations are built on exact diffraction theory, and leads to 
equations that can be solved numerically
with good accuracy without too much effort.

A number of calculations have addressed related 
problems.\cite{SNOM1,SNOM2,Alain,Take,Plasmth} 
One example of this is scanning near-field optical microscopy 
(SNOM),\cite{SNOM1,SNOM2,Pohl}
where one studies {\em photon} tunneling from a transparent sample to a tip. 
The tip-sample distance may be several hundred \AA\ in this case,
so it becomes necessary to take retardation into account.
But the calculations can often be carried out using various
discretization schemes in real space,\cite{Alain} 
since the  relevant length scales differ by at most an order of magnitude.
It deserves to be pointed out that the calculational scheme developed
in this paper could, with some extensions, also be applied to
SNOM problems, and may then be faster than the prevailing methods.
However, a prerequisite for this is that the tip-sample geometry
possesses some (preferably cylindrical) symmetry.

In another fully retarded calculation,
Takemori, Inoue, and Ohtaka\cite{Take} (TIO) studied the field around
a sphere in front of a plane surface with the aim of
investigating electromagnetic effects in surface-enhanced 
Raman scattering (SERS).\cite{MoskRMP}
Finally,
Madrazo, Nieto-Vesperinas, and Garcia\cite{Madra} (MNVG) have considered 
light emission from an STM in a calculation accounting for retardation.
They used an essentially two-dimensional (2D) model geometry, where
the tip was represented by a horizontal cylinder and the sample surface
was artificially corrugated.
We will compare our results with those of MNVG in as far as this is 
possible (they presented results for one single photon energy)
and we will see that
the different model geometries lead to quite different results.

The  inclusion of retardation effects does not change the basic picture
of how light is emitted from an STM, however, at a quantitative
level, there are changes to some of the calculated results.
As a general observation, retardation effects become increasingly 
important when the tip and/or sample materials are good conductors (such
as silver).
Then the energy dissipation in the near-field zone 
is rather small, and a localized mode will instead be damped because
some energy propagates away from the near-zone. This is of course allowed for 
in a retarded, but not in a non-retarded, calculation.
Thus for a Ag sample scanned by a Ir or W tip, 
retardation leads to an additional red-shift of the resonance
frequency, and this improves the agreement between theory and experiment.
For Au and Cu samples, on the other hand,
the resonance frequency is essentially unchanged, and there are
only relatively minor changes of the light emission intensity as 
a result of retardation effects.

The rest of the paper is organized as follows:
In Sec.\ \ref{Calcsec}, 
we develop the formalism needed to carry 
out a retarded calculation of the electromagnetic field that
results from inelastic tunneling.
We derive expressions for the radiated power, as well as the 
power dissipated in the sample and tip.
In Sec.\   \ref{Currsec}, 
we establish the connection between the 
completely classical current we use in the electrodynamic calculations,
and the transition matrix elements associated with inelastic 
tunneling of an electron from the tip to the sample or vice versa.
The numerical results of our calculation are presented in 
Sec.\ \ref{Ressec},
and compared with experimental results.
Finally, our conclusions are presented in Sec.\ \ref{Concsec}.

\section{The electromagnetic field calculation}
\label{Calcsec}

\subsection{General considerations}

We will calculate the 
electromagnetic field resulting from excitation 
by a {\em classical} current distribution
$\jvec(\rvec)e^{-i\omega t}$,
localized between the tip and the sample.
As in our previous, non-retarded calculations, 
we model the tip by a sphere with radius $R$ that is centered at the 
origin of our coordinate system, see Fig.\ \ref{Bild}.
The sample fills the half-space $z<z_0$, where
$z_0=-(d+R)$, $d$ being the tip-sample separation.

The optical properties of the tip and sample materials are described 
by local dielectric functions $\epsilon_1(\omega)$ (sample) and
$\epsilon_2(\omega)$ (tip) determined from experiment.\cite{Diel}
While this is an approximation, it is a reasonable one;
we will discuss its limitations in some more detail 
in Sec.\ \ref{Nonlocsec}.

We write the source current distribution as 
\be 
 \jvec(\rvec)=\hat{z}F(\rho)\delta(z-z_0)C.
\label{Modelcurr}
\ee
The source is  in other words a cylindrically symmetric dipole layer
right at the sample surface.
Most of the time we will take the radial form factor $F(\rho)$ 
to be a Gaussian
\be
 F(\rho)=e^{-\rho^2/a^2}
\label{Form}
\ee
with $a\sim $5 \AA\ (as long as the current is concentrated to a region near
the symmetry axis the exact shape of $F(\rho)$ is unimportant).
Even if we treat $\jvec$ as a classical current, its strength 
$C$, must eventually be determined from a 
quantum-mechanical calculation, 
which we defer to Sec.\ \ref{Currsec}.

Thanks to
the cylindrical symmetry of both the current distribution and
the tip and the sample, the magnetic flux density 
is always directed along the azimuthal unit vector $\hat{\phi}$, and
its magnitude only depends on $z$ and the distance $\rho$, to the 
cylindrical symmetry axis,
\be
 \Bvec(\rvec)=B(\rho,z)\hat{\phi}.
\ee
The accompanying $\Evec$ field  is also cylindrically symmetric.
In the region between the tip and the sample, it is mainly directed along
$\hat{z}$ and thus strongly coupled to $\jvec$.
Applying Maxwell's equations, we find that $\Bvec$ must satisfy the
vector Helmholtz equation 
(we use SI units and $\mu_0$ is the vacuum permeability) 
\be 
 \nabla^2\Bvec+k_i^2\Bvec=-\mu_0\nabla\times\jvec
\label{Helm}
\ee
in each of the three different regions of space. 
In vacuum
$k_0^2=k^2=\omega^2/c^2$, 
in the sample 
$k_{1}^2=\epsilon_{1}(\omega)k^2$, 
and in the sphere
$k_{2}^2=\epsilon_{2}(\omega)k^2$.
Inside the sphere the solution to Eq.\ (\ref{Helm}) can be written
\be 
 \Bvec=\hat{\phi} \sum_{l=1}^{\infty} a_l \frac{j_l(k_2 r)}{j_l(k_2 R)}
 P_l^1(\cos{\theta}), 
\label{Bsphere}
\ee
where $r$, $\theta$, and $\phi$ are the usual spherical coordinates,
$j_l$ is a spherical Bessel function, $P_l^1$ is an associated Legendre 
function, and $a_l$ is a set of coefficients that remain to be determined.
In the sample we write $\Bvec$ as a Fourier-Bessel transform,
\barr
 \Bvec=\hat{\phi}\int_0^{\infty}&& \kappa d\kappa \ 
  B(\kappa) J_1(\kappa \rho)
 \breakeq
 \times
  \exp{\left[-i\sqrt{k_1^2-\kappa^2}\ (z-z_0)\right]},
\label{Bsamp}
\earr
where  $J_1$ is an ordinary Bessel function, while $B(\kappa)$ is an 
as yet unknown function.
Equation (\ref{Bsphere}) yields a finite field at the 
center of the sphere, 
Eq.\ (\ref{Bsamp}) gives outgoing, damped waves in the sample,
and both expressions satisfy Eq.\ (\ref{Helm}).
Next we must determine $a_l$ and $B(\kappa)$.

\subsection{The solution}
\label{Solu}

To calculate the field in the vacuum region, we use the 
vector equivalent of the  Kirchoff integral\cite{Jack}
\barr
 &&\Bvec(\rvec)=\Bvec^{(e)}(\rvec)+
\breakeq
 +\int dS' [(\hat{n}'\times\Bvec(\rpvec))\times\nabla'G
   -\frac{ik}{c}G(\hat{n}'\times\Evec(\rpvec))].
\label{Kirchoff}
\earr
Here, the first term $\Bvec^{(e)}$ solves Eq.\
(\ref{Helm}) when neither the sample nor the tip is present,
i.e., it yields the direct contribution to $\Bvec$ from the source.
$G$ stands for the Green's function to the scalar 
Helmholtz equation in vacuum,
\be
 G(\rvec,\rpvec)=\frac{e^{ik|\rvec-\rpvec|}}{4\pi|\rvec-\rpvec|}.
\label{Greendef}
\ee
The integration $dS'$ in Eq.\ (\ref{Kirchoff})  runs over two surfaces:
a plane just outside the sample and 
a spherical surface just outside the model tip.
The normal vector $\hat{n}'$ points in
the direction from the sphere or half-space into the vacuum region.
The fields $\Bvec(\rpvec)$ and $\Evec(\rpvec)$ are the exact fields on
the two surfaces, so Eq.\ (\ref{Kirchoff}) is an integral equation.

Before continuing, let us point out
that even if the surface integral in Eq.\ (\ref{Kirchoff})
contains the full $\Bvec$ and $\Evec$ fields on the surfaces,
the combination appearing in the integral guarantees that the
final result only contains  reflected waves.
If one for example wants to calculate the field from a certain 
source distribution in vacuum, but nevertheless introduces a surface
somewhere outside the source, the integral 
does not at all contribute to $\Bvec(\rvec)$.
In the present case,
the integral over the spherical surface vanishes if $\epsilon_2=1$,
and the integral over the flat surface vanishes if $\epsilon_1=1$.
 
Since Eq.\ (\ref{Kirchoff}) yields the field $\Bvec$ at any point 
outside the tip and sample, it can be used to find expressions
determining $a_l$ and $B(\kappa)$. 
Just outside the surfaces over which we integrate,
we can (i) either calculate the field from Eq.\ (\ref{Kirchoff}), or
(ii) we can use Eq.\ (\ref{Bsphere}) or Eq.\ (\ref{Bsamp})
because $\Bvec$ is continuous across the interfaces.
The so obtained, different expressions for $\Bvec$ must be equal.
Expanding the result outside the sphere in terms of associated 
Legendre functions yields
\be
 a_l=a_l^{(e)}+s_la_l+\int_0^{\infty} \kappa d\kappa\ f_l(\kappa)B(\kappa),
\label{aleq}
\ee
where $a_l^{(e)}$ comes directly from the source, the term
$s_la_l$ comes from the integration over the sphere, and the last term
originates from the sample surface integration.
Without the sample, the above equation has the solution
$a_l=a_l^{(e)}/(1-s_l)$; 
thus, plasmon resonances occur on the isolated sphere when $s_l=1$.
The corresponding equation, valid at the sample surface, is
\be
 B(\kappa)=B^{(e)}(\kappa)+S(\kappa)B(\kappa)+\sum_{l=1}^{\infty}
 a_l g_l(\kappa).
\label{Bkeq}
\ee
Removing the sphere, $B(\kappa)=B^{(e)}(\kappa)/(1-S(\kappa))$,
so the surface plasmon resonances occur when $S(\kappa)=1$.
The explicit expressions for $B^{(e)}(\kappa)$, $f_l(\kappa)$, $g_l(\kappa)$,
$s_l$, $S(\kappa)$, $a_l^{(e)}$, and $a_l^{(p)}$ 
are given below and in the Appendix.

To solve for $a_l$ and $B(\kappa)$, we first note that Eq.\ (\ref{Bkeq})
has the formal solution
\be
 B(\kappa)=
 \frac{B^{(e)}(\kappa)+\sum_{l=1}^{\infty} a_l g_l(\kappa)}
 {1-S(\kappa)}.
\label{Bksol}
\ee
Inserted into Eq.\ (\ref{aleq})
this yields a system of linear equations determining the coefficients $a_l$, 
\be
 [1-s_l]a_l
  -\sum_{l'=1}^{\infty}M_{ll'}a_{l'}
  =a_l^{(e)}+a_l^{(p)},
\label{Mateq}
\ee
where 
\be
 M_{ll'} =\int_0^{\infty} \kappa d\kappa\  
 \frac{f_l(\kappa)g_{l'}(\kappa)}{1-S(\kappa)}.
\label{Mllint}
\ee
The terms on the right hand side are the driving forces and depend
on the source: 
$a_l^{(e)}$ is the direct contribution from the source,
while $a_l^{(p)}$ results from the source fields reflected {\em once} off
the sample surface.
The two coefficients can be added together to give
\be
 a_l^{(e)}+a_l^{(p)}=\int_0^{\infty} \kappa d\kappa\ 
 \frac{f_l(\kappa)B^{(e)}(\kappa)}{S(\kappa)[1-S(\kappa)]}.
\ee
$B^{(e)}(\kappa)$ is the Fourier-Bessel transform of the
solution to Eq.\ (\ref{Helm}) at $z=z_0$ and with $\epsilon_1=1$ and
$\epsilon_2=1$.
Using the expression in Eq.\ (\ref{Gcyl}) for the 
Green's function, we get
\barr 
 B^{(e)}(\kappa)&&=\frac{-i\mu_0 C}{2\sqrt{k^2-\kappa^2}}
 \int_0^{\infty} \rho d\rho\ F'(\rho)J_1(\kappa\rho)=
\breakeq
 =\frac{i \mu_0 C}{\sqrt{k^2-\kappa^2}}
   \frac{\kappa a^2}{4}\ e^{-\kappa^2 a^2/4},
\earr
where the second line is valid if $F(\rho)$ is given by
Eq.\ (\ref{Form}).

To summarize,
a numerical calculation of the electromagnetic field
begins by evaluating
$s_l$, $M_{ll'}$, and 
$a_l^{(e)}+a_l^{(p)}$ from the formulas given here and in the Appendix.
Then the coefficients $a_l$ can be found by a matrix inversion in 
Eq.\ (\ref{Mateq}). 
Having found $a_l$, we get $B(\kappa)$ from Eq.\ (\ref{Bksol}),
and subsequently $\Bvec(\rvec)$ can be calculated anywhere in space.
These steps must of course be repeated for each photon frequency 
$\omega$.

To carry out the matrix inversion in practice,
a truncation in $l$ at $l_{\rm max}$
must be introduced.
With the geometry parameters that we consider most of the
time ($d\sim$ 5\AA, and $R\alt $ 500 \AA), $l_{\rm max}\sim 50$ is
appropriate in order to get converged results.
$l_{\rm max}$ is determined by the size of the resonant
cavity between the tip and the sample.
The interface-plasmon mode is to a large extent confined within the region
where the distance between tip and sample is less than twice the
{\it smallest} tip-sample separation $d$ (cf.\ Ref.\ \onlinecite{Rend}).
For a spherical tip this region has a radius
$\sqrt{2dR}$ (if $d\ll R$), i.e., with $d=5$ \AA\ and $R=300$ \AA\ we get a 
``plasmon radius'' of $\approx55$ \AA\@. 
Since $\pi R\approx 950$ \AA, it is clear that even a crude description 
of the interface plasmon would require $l_{\rm max}\sim 20$, but to capture
its structure a larger number of basis functions is needed.

The computation of the radiated power for one frequency takes
less than 5 seconds on an ordinary work station.
Most of this time is used for the $\kappa$ integration 
determining $M_{ll'}$ in Eq.\ (\ref{Mllint}).

\subsection{Emitted power}

Once the coefficients $a_l$ have been determined, we can evaluate 
the radiation power as well as the power dissipated into the sample and tip.
In vacuum, far from the model tip, 
the $\Bvec$ field has the asymptotic behavior
\be
 \Bvec=\hat{\phi}\bfar\frac{e^{ikr}}{r}.
\ee
$\bfar$ is a sum of contributions from the source, the sample surface,
and the spherical surface,
\barr
 \bfar&&
 =-
 [1+S(k\sin{\theta})]
 k\cos{\theta} B(k\sin{\theta})
 e^{ik(d+R)\cos{\theta}}
\breakeq
 -\sum_{l=1}^{\infty} 
  a_l \frac{e^{-il\pi/2}}{k}
 \sqrt{\frac{2l+1}{4\pi}}
  P_l^1(\cos{\theta})
 \frac{s_l}{h_l(kR)}.
\earr
Knowing $\bfar$, we can calculate the radiated (differential) 
power per unit solid angle
\be 
 \frac{dP_{rad}}{d\Omega}=\frac{1}{2}c^3\epsilon_0|\bfar|^2.
\label{Prad}
\ee

By evaluating the Poynting vector on the surface of the 
model tip and then taking the time average, we find that
the power dissipated into the sphere is given by
\barr
&& P_{\rm sph}=\pi R^2\frac{c^3\epsilon_0}{k}
\breakeq
\times
 \sum_{l=1}^{\infty} \frac{2l(l+1)}{2l+1}|a_l|^2\ {\rm Im}
\left[-\frac{1}{\epsilon_2}
 (\frac{1}{R}+k_2\frac{j_l'(k_2R)}{j_l(k_2R)})\right] .
\label{Psph}
\earr
A similar calculation yields
the power absorbed in the sample
\barr
 P_{\rm samp}=\pi \frac{c^3\epsilon_0}{k}
 \int_0^{\infty} \kappa d\kappa |B(\kappa)|^2
  \ {\rm Re}\left[\frac{\sqrt{k_1^2-\kappa^2}}{\epsilon_1}\right] .
\label{Psamp}
\earr

\subsection{Comparison with other methods}
\label{Compare}

Takemori, Inoue, and Ohtaka\cite{Take} 
studied scattering of a plane wave impinging on a sphere-plane system.
This problem can be studied also in the present framework
if $\Bvec^{(e)}$ is taken to describe the incoming wave.
Of course, the solution would not in general be restricted to 
cylindrically symmetric electric multipole modes as is the case here,
but thanks to the cylindrical symmetry there would not be any mixing of
modes with different $m$. The numerical solution would therefore be quite 
feasible.
Note also that thanks to the reciprocity theorem,\cite{Reci} already
the present calculation can yield results for the $z$ component of the 
electric field induced by an incoming wave
on the cylindrical symmetry axis below the tip. 

The conversion of plane waves to spherical waves and vice versa
is central to both our solution and that of TIO\@. 
However, the conversion is dealt with in different ways. 
TIO essentially sum up the contributions from repeated scattering,
back and forth, between the sphere  and the plane.
These repeated scattering events are implicitly included in the 
Kirchoff integral formulation. Thus, we feel that, at least once
the field has been written in the form of Eq.\ (\ref{Kirchoff}),
the present solution is conceptually simpler.

Our calculation and the one by
Madrazo, Nieto-Vesperinas, and Garcia\cite{Madra,MadraJOSA}
have similar starting points; 
in both cases the electromagnetic field is expressed in terms of
integrals over the surfaces where the relative dielectric function
changes in a discontinuous way.
However, unlike us, MNVG choose to solve the resulting integral
equations in real space.
While MNVG treat a situation where the sample surface is periodically
corrugated, it seems like their method would require quite 
intensive numerical calculations even for the case of a flat sample.

\section{The source}
\label{Currsec}

In order to perform the explicit calculations, we need to specify the
sources.
The tunnel current of an STM is typically concentrated to a small area; 
most of the current goes through the very last atom at the end of the tip.
Also the inelastic transitions, from a state in one 
of the electrodes to a lower-energy state in the other electrode, 
that drive the light emission are concentrated to the same small region 
of space.

We wish to relate the classical ac current $\jvec$ in 
Eq.\ (\ref{Modelcurr}) to quantum-mechanical transition matrix elements.
Let us first look at one particular electronic transition from the 
state $|i\rangle$
to $|f\rangle$.
The corresponding classical current is\cite{Blatt}
\be
 \jvec^{(fi)}(\rvec)=\langle f|2 {\bf j}(\rvec)|i\rangle.
\ee
The resulting current distribution is of course not identical to that
of  Eqs.\ (\ref{Modelcurr}) and (\ref{Form}), however, 
the current is mainly directed along $\hat{z}$ and
it is concentrated to a small region near the end of the tip. 
As a consequence, the two current distributions generate essentially the
same fields provided their dipole moments
$$ 
 \int d^3r z  \rho_e(\rvec) {\rm \ \ and \ \ }
 \int d^3r z  \rho_e^{(fi)}(\rvec)
$$
are equal.
Here the charge distributions $\rho_e$ and $\rho_e^{(fi)}$ are connected 
to $\jvec$ and $\jvec^{(fi)}$ through
continuity equations such as
\be
 \frac{\partial \rho_e}{\partial t}+\nabla\cdot\jvec=0.
\ee
In order for the dipole moments to be equal,
we must have
\be
 -\frac{2\pi C_{fi}}{i\omega}\int_{0}^{\infty}
  \rho d\rho F(\rho)
  =-\frac{1}{i\omega}\int d^3r \langle f|2j_z(\rvec)|i\rangle.
\label{Cfidef}
\ee
Here $C_{fi}$ is the value $C$ would take if 
$(|i\rangle \rightarrow |f\rangle)$
was the only inelastic transition contributing to photon emission.

In the experiment, light emission is caused by 
many different transitions at a whole range of frequencies.
Therefore the factor $C$ entering Eq.\ (\ref{Modelcurr}) should be an 
{\em incoherent sum} of the different $C_{fi}$;
moreover, we should calculate the emitted power {\em per unit photon energy}
rather than just the emitted power.
All the expressions for radiated or dissipated power  found from
(Eqs.\ (\ref{Prad}), (\ref{Psph}), and (\ref{Psamp}))
are proportional to $|C|^2$.
We need to replace $|C|^2$ by a new quantity, 
 $|{\cal C}|^2(\omega)$,
in order to get expressions for the radiated or dissipated power 
per unit photon energy.
For the  transitions around photon energy $\hbar \omega$, we have
\be
 |{\cal C}|^2(\omega)
 =\sum_{fi} |C_{fi}|^2 \delta(E_i-E_f-\hbar \omega).
\ee
With the aid of Eq.\ (\ref{Cfidef}),
this can be rewritten as 
\be
 |{\cal C}|^2(\omega)=
 \sum_{fi} \frac{|\langle f |2j_z|i\rangle|^2}{4\pi^2
 |\int_0^{\infty} \rho d\rho F(\rho)|^2}\delta(E_i-E_f-\hbar \omega).
\ee
Now, for example, 
the radiated power per unit solid angle and photon energy is 
\be 
 \frac{dP_{rad}}{d(\hbar\omega)d\Omega}=
 \frac{|{\cal C}|^2(\omega)}{|C|^2}
 \ \frac{1}{2}
 c^3\epsilon_0|\bfar|^2
\ee
if the calculation of $\bfar$ is still based on a source described
by Eq.\ (\ref{Modelcurr}).

In the actual calculation of $|{\cal C}|^2(\omega)$,
we have used free-electron models for both the tip and sample;
more details can be found in Ref.\ \onlinecite{ZPB}.
The resulting $|{\cal C}|^2$ is rather featureless;
the $\omega$ dependence is roughly
\be
 |{\cal C}|^2(\omega)\propto
  \left(1-\frac{\hbar\omega}{eV_{\rm bias}}\right),
\label{powspec}
\ee
where $V_{\rm bias}$ is the bias voltage.
We have not tried to improve on these earlier calculations, for
the effects we are mainly interested in are of 
electrodynamic origin.\cite{Tsuk}

We should point out that we restricted the evaluation of the 
transition matrix elements to the vacuum (barrier) region.
There are several good reasons for doing that:
Since, in the frequency range where the photon emission is most intense,
$|\epsilon_{\rm tip}|$ and $|\epsilon_{\rm sample}|$ are considerably
larger than 1, the electric field that couples to the tunnel current
is stronger in the vacuum region. 
Moreover,
the electron wave functions oscillate inside the sample and tip,
whereas they do not change sign in the barrier region.
This means that the vacuum region should give the largest 
contributions to the matrix elements.

Persson and Baratoff\cite{BNJP} studied this issue rather thoroughly
by comparing the contributions to photon emission from inelastic tunneling
and hot-electron decay. 
Within their model, in which electrons are tunneling 
into a spherical metal particle, 
inelastic tunneling processes (light emission while the electron is 
in the barrier) are about
3 orders of magnitude more effective than hot-electron decay
(light emission when the electron has reached the final-state electrode)
in the light-emission process. 

\section{Results}
\label{Ressec}

We have calculated the electromagnetic field for a number of different 
materials combinations and geometric parameters.
Below we present results for the differential power
(i.e.\ the radiated power per unit solid angle and photon energy)
and
the total radiated or dissipated power per unit photon energy.

When we present results for the differential power,
we take the observation angle to be $\theta=57.3^\circ$ (i.e.\ 1 rad).
The angular distribution of radiation shows only slight variations
with frequency, materials, etc. 
It is given by a dipolar radiation lobe that 
has been ``turned'' away from $\theta$=90$^\circ$ to have a maximum
at $\theta$=55--60$^\circ$. 
Very similar angular distributions show up both
in experiments (Fig.\ 30 of Ref.\ \onlinecite{RBThes})
and other calculations (see Refs.\ \onlinecite{Ueh} and \onlinecite{Take}).

\subsection{Au sample---W tip}

We begin by considering light emission from a Au sample probed by a
W tip. Figure \ref{AuWfig} presents results for the differential 
power from both retarded and non-retarded calculations.
As for the spectral shape, these results should be directly 
comparable with experiment (typically photons are collected over 
a certain solid angle in the experiments).\cite{PRL91}
Here, retardation does not cause any qualitative changes of the spectra.
For a tip radius $R$=100 \AA, the retarded and 
non-retarded results cannot be distinguished from each other.
When $R$ is increased to 300 \AA, there is a  difference in terms of 
intensity between the retarded and non-retarded results,
but the spectral shape and resonance frequency is nearly the same.
These spectra are in quite good agreement with the 
experimental ones.\cite{PRL91}

When the sample consists of Au (and the same holds true for Cu), 
the resonance frequency 
is not very sensitive to the tip-sample geometry.
The reason is that, because of the onset of interband
transitions, the real part of the dielectric function $\epsilon_{\rm Au}$ 
increases very rapidly from $\approx$-11 at 2 eV 
to $\approx$-2 at 2.5 eV\@.
Thus, even though  changes in the geometry leads to changes 
in the resonance condition expressed in terms of a value for the 
sample dielectric function, the resonance {\em frequency} does not change
much; here there is only a slight red-shift of the resonance when $R$
increases from 100 to 300 \AA\@.

Umeno {\it et al}.\cite{Umeno} observed light emission from a 
granular Au film, and found variations in the resonance frequency
that they interpreted as the result of a varying tip curvature.
That interpretation is in qualitative agreement with our results,
but a quantitative comparison is not possible for at least two reasons:
(i) In its present form, 
our theory does not account for the granularity of the film. 
(ii) The bias voltage was as low as 2 V in the experiment,
so the resonance frequency was most likely limited by $V_{\rm bias}$. 
It would be interesting to see the experiment repeated at a larger bias.

\subsection{Ag sample---Ir tip}

Next let us look at the radiation spectra from a silver sample 
scanned by an Ir tip.    In  Fig.\ \ref{AgIrfig}
we display results of calculations including retardation with
three different tip radii (200, 300, and 400 \AA, respectively),
and for comparison also results from a non-retarded 
calculation with $R$=300 \AA\@.
Here, the retarded and non-retarded results 
still have the same basic features,
but the inclusion of retardation effects leads to substantial changes
at the quantitative level.
The primary resonance frequency is red-shifted, 
the second peak (between 3 and 3.5 eV) reduces to a shoulder,
the peak height is reduced by almost a factor of 2, 
and the peak becomes broader.

One important reason for the lowering of the resonance frequency and
increased damping (width) of the resonance appears to be that
the fields penetrate further into the sample and tip when 
retardation effects are included in the calculation.
This increases the resistive losses suffered by the tip-induced 
plasmon mode. 
We would like to stress that radiation damping as such plays
{\em no} role in broadening the resonance here;
as we will see below, the radiated power is  just a small 
fraction of the dissipated power.

The changes brought about by the retardation effects
improve the agreement between calculated and experimental spectra.
The second peak, seen in the non-retarded calculation but not in experiment,
is gone, and the extra red-shift found in the retarded calculation
brings the remaining resonance closer to its experimental position
(typically 2.5 eV).\cite{EPL}
At the same time, a word of caution is in place here. 
The calculated spectra are clearly quite sensitive to the tip-sample geometry.
In experiments, the peak positions varied somewhat 
between spectra taken with different tips on silver samples.\cite{PRL91}
However, these variations were not as large as those 
seen in Fig.\ \ref{AgIrfig}, where the resonance frequency is
lowered by $\approx$0.5 eV when $R$ is increased from 200 \AA\ to 400 \AA.
Unfortunately, our limited knowledge about the actual shape of the STM tip
makes it difficult to reach any definite conclusion about this issue.

Further insight can be gained by studying how 
the total radiated power, as well as the power lost to the
tip and sample, vary with frequency. Such results are displayed in 
Fig.\ \ref{Totpowfig} for the case of a Ag sample 
and an Ir tip with $R$=300 \AA.
The losses to the tip and sample are much larger than the 
radiated power (note that the $P_{\rm rad}$ data are multiplied by 
a factor of 100). 
From the figure, the {\em total} radiated power can be estimated to
$$
25\times10^6 \frac{\rm W}{\rm J}\times 1 {\rm eV}\approx 4 pW,
$$
or 10$^{7}$ photons/sec.
The tunnel current 10 nA corresponds to 
$\approx 0.62\times10^{11}$ tunneling electrons per second.
Thus, the quantum efficiency is about
2$\times$10$^{-4}$ photons/tunneling electron. This is 
in reasonable agreement both with the results of our earlier 
calculations\cite{JMA,ZPB} and experimental estimates.\cite{EPL,PRL91}

\begin{figure}
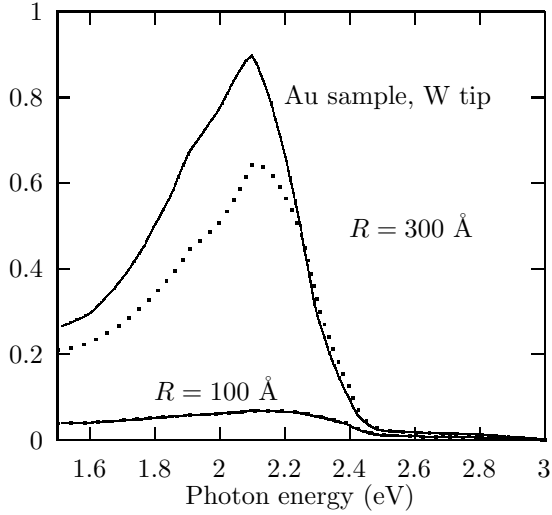

\medskip
 \caption{Radiated power per unit solid angle and photon energy
for the case of a gold sample and tungsten tip.
The bias voltage is 3 V (negative sample) and the tunnel current is 10 nA.
The corresponding tip-sample separation $d$ was 5.71 \AA\ ($R$=100 \AA) and 
6.30 \AA\ ($R$=300 \AA). The continuous curves show results of retarded 
calculations, while the dotted curves show results calculated without 
accounting for retardation.}
 \label{AuWfig}
\end{figure}

\begin{figure}
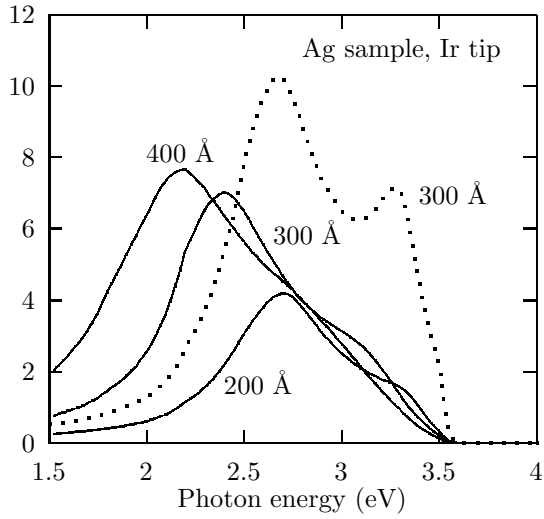

\medskip
 \caption{
Radiated power per unit solid angle and photon energy
with a silver sample and iridium tip. 
The bias voltage is 4 V, the current is 10 nA, 
and the corresponding tip-sample separations are: 
$d$=6.09 \AA\ (for $R$=200 \AA), 
$d$=6.28 \AA\ ($R$=300 \AA), 
and
$d$=6.41 \AA\ ($R$=400 \AA). 
As in Fig.\ \protect\ref{AuWfig}, the continuous (dotted) 
curve shows results from a retarded (non-retarded) calculation.
}
 \label{AgIrfig}
\end{figure}

\begin{figure}
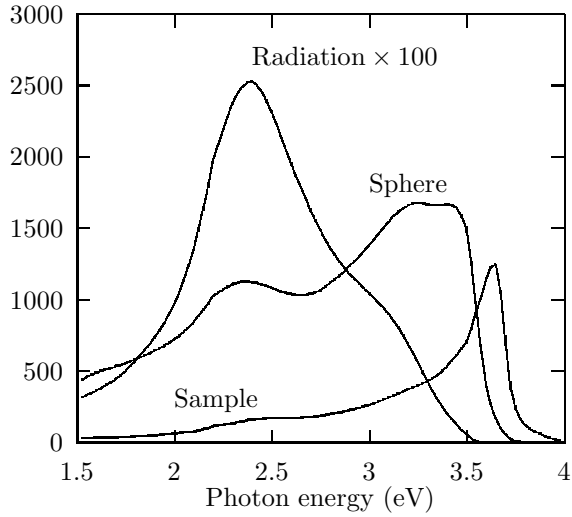

\medskip
 \caption{The total power per unit photon energy
 emitted as radiation, and dissipated into the tip and sample.
 The calculation takes retardation into account.
 The tip and sample materials are Ir and Ag, respectively, 
the tip radius $R$=300 \AA,
and the rest of the parameter values are the same as in Fig.\
 \protect\ref{AgIrfig}. To facilitate direct comparison, the radiation power 
data have been multiplied by a factor of 100.}
 \label{Totpowfig}
\end{figure}

Here one should
note that the electromagnetic response is not always
the limiting factor for the quantum efficiency.
To reach the values found above, the bias voltage must be 1 V, or so,
larger than the resonance photon energy.
But if the bias voltage is just 2 V, and in addition
a Au sample and a W tip 
is used, the quantum efficiency drops to 10$^{-6}$ or less
(this is typically what was found in the experiments in
Refs.\ \onlinecite{Sivel} and \onlinecite{Umeno})\@.
This estimate is found by noting that: (i) if the bias voltage is 2 V,
in view of Eq.\ (\ref{powspec}), less than 10 \% of the area 
under the curves in Fig.\ \ref{AuWfig} remains, and (ii)
comparing Figs.\ \ref{AuWfig} and \ref{AgIrfig} one realizes 
that the quantum efficiency for a Au-W configuration 
is about an order of magnitude smaller than for a Ag-Ir configuration.

Returning to Fig.\ \ref{Totpowfig} we see that the dissipated power
is about 100 times as large as the radiated power;
thus, since about 2 out of 10$^4$ tunneling electrons 
take part in photon emission
processes, it is clear that at most a few percent
of all the electrons undergo some
inelastic process while in the gap between the tip and sample.
(The calculation of Ref.\ \onlinecite{Downes}, using a two-sphere
model geometry, showed that the inelastic current could account for 
up to 10 \% of the total current.)
In any case, the total tunnel current is dominated by the elastic part
and the channels opened by inelastic processes only account for a
rather small correction to the current.
Most of the energy supplied by the bias voltage is eventually
dissipated in
hot-electron decay processes well inside the tip or sample.

These observations are important because 
our scheme for determining the tip-sample distance $d$
that gives a certain tunnel current $I$ at a certain 
bias voltage $V_{\rm bias}$ is now justified.
In that calculation (see Ref.\ \onlinecite{ZPB}) 
we take $I=I_{\rm elastic}$.
{\it A priori}, nothing guarantees that the elastic contributions to
the tunnel current are the dominating ones; 
given a sufficiently strong coupling to some other degrees of freedom, 
inelastic tunneling could dominate.

In the present case, the resistive losses into the tip (Ir) dominate 
for most frequencies; 
Ir is a rather bad conductor.
However, just above 3.5 eV the losses into the sample are larger.
This is mainly due to surface plasmon
emission (the surface plasma frequency of silver is $\approx$ 3.7 eV).
Then the tunneling electron excites a surface plasmon that can propagate
quite far away from the tip-sample cavity, but eventually loses 
its energy due to dissipation into the sample.

\begin{figure}
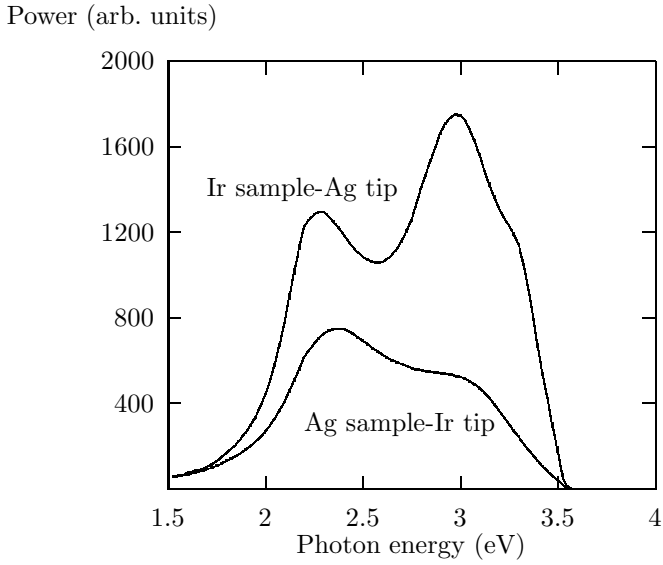

\medskip
 \caption{
 Effects of interchanging tip and sample materials.
 The two curves show the calculated differential power, given
 that the driving current has a white spectrum.\protect\cite{white}
}
\label{Swapfig}
\end{figure}

Let us finally see what happens  when {\em the tip and sample materials
are interchanged}. 
Figure \ref{Swapfig} compares the emission rates for a Ag sample-Ir tip 
with that of the combination Ir sample and Ag tip.
In this calculation we used a ``white'' spectrum for the 
source.\cite{white}
As is seen, the magnitude of the emitted power can change
by a factor of 2--3 as a result of the interchange in our model.
Considering the details of the spectra there are further changes.
The spectrum obtained with a Ag model tip has two peaks, and the one
at higher frequency has the largest magnitude.
Clearly, this peak originates, at least in part, from the fact that the 
model tip is a sphere, since it appears not very far below the frequency
where an isolated silver sphere has a resonance.

\subsection{Ag sample---Ag tip}

The effects of retardation become even more apparent when we 
consider a situation in which both the tip and the sample are made
of silver. 
Experiments have been done with tungsten tips covered by silver.\cite{PRL91}
The obtained spectra typically had two resonances.
However, since the exact ``composition'' of the tip in these experiments
is unknown,
we will not calculate absolute intensities or make any comparison
with experiment, but rather look at trends and make relative 
comparisons.

\begin{figure}
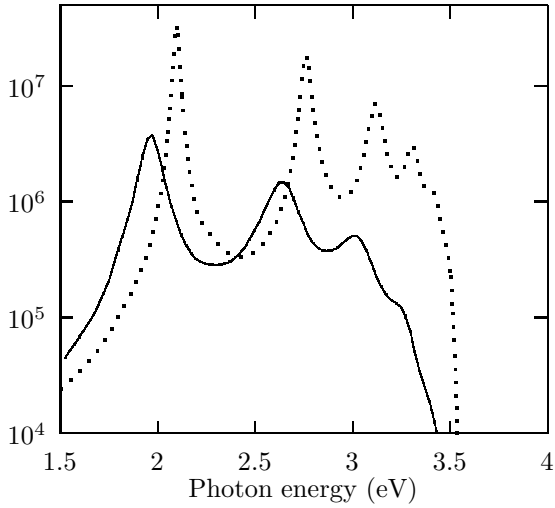

\medskip
 \caption{Radiation amplification factor for an Ag sample  and Ag tip
with $R$=300 \AA\ and  $d$=5 \AA.
The amplification factor is the ratio between the radiated differential 
power with tip and sample present, and the radiated power into vacuum 
given the same source distribution. 
The dotted curve gives the results from a non-retarded calculation.
The peaks of this curve are about 10 times 
higher and sharper than those of the continuous 
curve resulting from a retarded calculation. Note the logarithmic scale.
}
 \label{AgAgfig}
\end{figure}

In Fig.\ \ref{AgAgfig}, we display results for the radiation amplification
factor, i.e.\ the actual differential power 
divided by the value it would take if the tip as well as
the sample was absent, 
comparing the non-retarded and retarded calculations.
Silver is a very good conductor over the entire 
frequency range up to $\hbar\omega$=3.5 eV, and tip-induced 
plasmons localized to the region around the tip-sample gap
are very weakly damped, unless energy can be dissipated due to wave 
propagation. Therefore the spectrum resulting from
a non-retarded calculation has a series of 
very sharp peaks.
Once the finite speed of light is taken into account, these resonance peaks 
are broadened and their heights are reduced by about an order of
magnitude. Still the amplification  (enhancement) is very high.
For example, if the corresponding curve was plotted for a Ag sample 
and an Ir tip, the peak value would be $\approx$10$^5$. 
The changes  brought about by including retardation
can be understood by essentially 
the same reasoning as we used before.

\begin{figure}
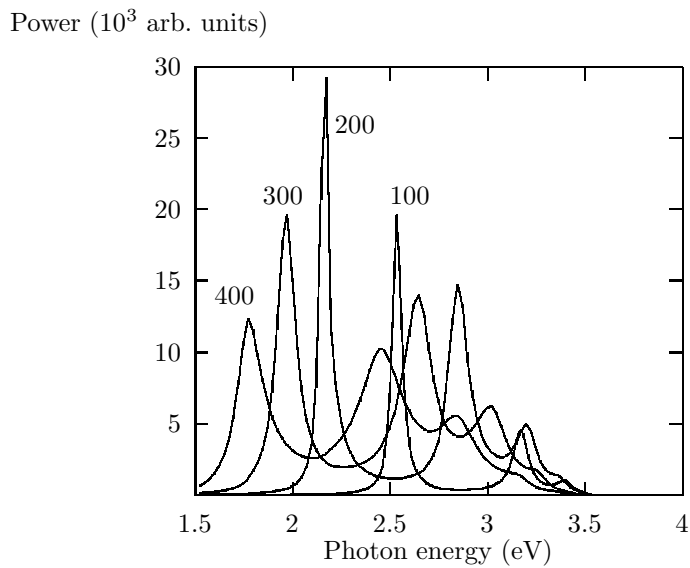

\caption{
Results for the  differential power
for the case of a Ag sample and Ag tip. 
A white power spectrum\protect\cite{white} was used for the current 
driving the light emission.
The calculations were done taking retardation into account.
The tip radii were  100, 200, 300, and 400 \AA, respectively, 
as indicated next to the curves, and $d$=5 \AA.
}
\label{AgAgRfig}
\end{figure}

In Fig.\ \ref{AgAgRfig}, we show how the differential power 
(calculated with a white power spectrum for the current\cite{white}
develops with increasing tip radius. 
Each spectrum has several, rather sharp peaks.
The different resonances correspond to modes with different
field distribution in the gap between the tip and the sample.
The resonance with the lowest frequency is nodeless, the second peak
comes from a mode that has one node, etc.
As before, the resonances are red-shifted and broadened when $R$ increases.
As for the peak height, it first grows because a larger 
sphere functions as a larger ``antenna'' putting out more power.
However, already at $R$=300 \AA\ the peaks are lower and broader
due to the increased damping. 
When this happens, the total
radiated power remains approximately constant to begin with.
But eventually also this trend is broken once the size of
the sphere becomes comparable to $\lambdabar$ ($\lambda/2\pi$).
Then different parts of the sphere cannot send out radiation coherently
any longer.

\begin{figure}
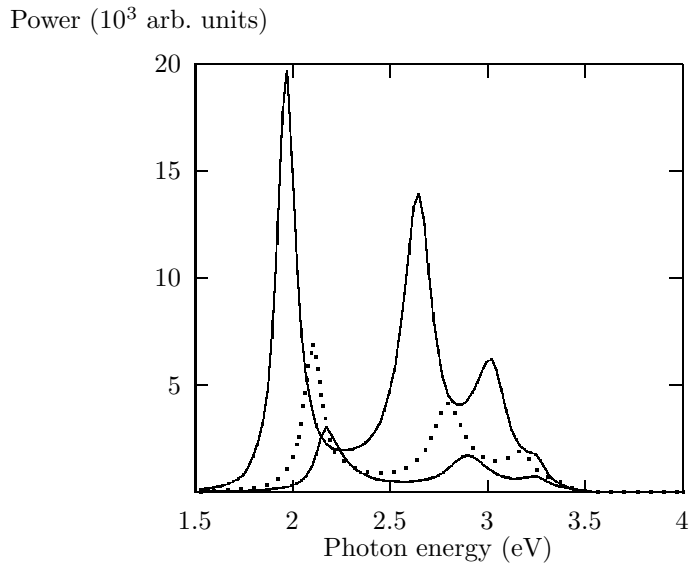

\caption{
Differential power 
(calculated with a white spectrum\protect\cite{white} 
for the driving current) for a Ag sample and 
Ag tip with $R$=300 \AA\ 
for varying tip-sample distance.
$d$=5 \AA\ for the curve with highest peaks,
$d$=7.5 \AA\ for the dotted curve, and
$d$=10 \AA\ for the curve showing the lowest intensity.
}
\label{dfig}
\end{figure}

The differential power
spectra are of course also sensitive 
to the tip-sample separation $d$. 
This is illustrated in Fig.\ \ref{dfig}.
Keeping $R$ fixed and increasing
$d$, the resonances are shifted upwards in frequency due to
a weaker coupling between the tip and sample.
At the same time, the peak values
decrease; with a larger tip-sample separation the field enhancement 
in the cavity becomes less effective, and the total photon yield 
decreases monotonically with increasing $d$.
If one just looks at the intensity at a certain fixed photon energy,
however, it can happen that the light intensity first increases as 
the tip approaches the surface, but then decreases again.
This is due to the shift of the resonances as the tip-sample
distance is changed; for a certain $d$ a resonance appears at the 
frequency in question. 
An example of this is seen around $\hbar\omega=2.1$ eV in 
Fig.\ \ref{dfig}.

\begin{figure}
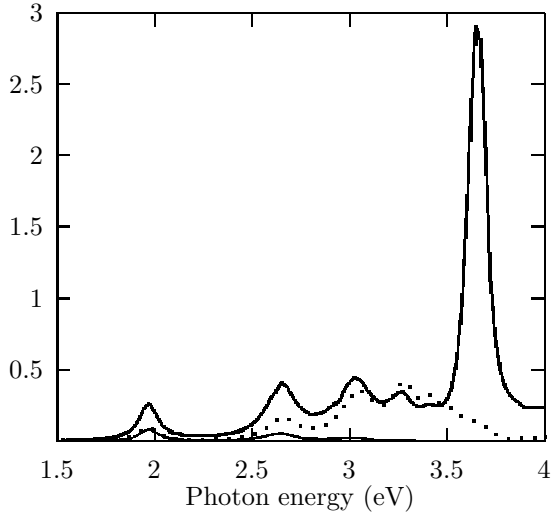

\caption{Total radiated, and dissipated 
power per unit photon energy
for a Ag sample and Ag tip (R=300 \AA\ and d=5 \AA). 
The thin curve gives the total radiated power, 
the other curves show the dissipated power into the 
sample (thick curve) and tip (dotted curve).
The results were calculated with
a white spectrum\protect\cite{white} for the current.
}
\label{AgAgpowfig}
\end{figure}

In Fig.\ \ref{AgAgpowfig}, we have plotted the total radiated and dissipated
power calculated with a white power spectrum\cite{white} for the current.
There is a  conspicuous peak in $P_{\rm samp}$
just above 3.5 eV,
which is due to surface-plasmon excitation.
However, over most of the rest of the frequency range, the radiated 
power is comparable to $P_{\rm samp}$ and $P_{\rm sph}$.
Thus, in the present case, radiation damping really gives a significant
contribution to the peak width.

\subsection{Surface modifications to the dielectric response}
\label{Nonlocsec}

As mentioned earlier, using dielectric functions determined from 
optical experiments is an approximation 
(however, as we will see, a good one)
in the present problem.
A more accurate dielectric function should include effects of extra 
damping due to surface scattering and electron-hole pair excitations.
Here we will try to give an idea about the size of this damping 
using rather simple estimates.

To estimate the change in $\epsilon$ due to surface scattering, 
let us assume that the
dielectric function at least over a certain frequency range
can be written on the Drude form
\be
 \epsilon(\omega)=1-(\frac{\omega_p}{\omega})^2
 + i\frac{(\omega_p)^2}{\omega^3\tau},
\ee
where $\omega_p$ is the plasma frequency, set by the 
density of conduction electrons, and $\tau$ is the electron mean free
time.
A simple estimate of $\tau$ is just $\tau=\ell/v_F$, where
$\ell$ is a mean free path and $v_F$ the Fermi velocity.
Extra scattering off the surface should primarily occur in the tip,
because it has of course a different shape than the flat surfaces
used in the optical measurements in which the local dielectric 
functions were determined.
But the standard tip materials Ir and W already have
dielectric functions with large imaginary parts so an extra 
damping mechanism there makes no big difference.
Let us therefore look at Ag, for which $\omega_p\approx 9 eV$,\cite{Agplasm}
and $v_F\approx 1.4\times10^6$ m/s. 
With $\ell=300$ \AA\ (i.e.\ the tip radius we used most often)
and $\hbar\omega=2.5$ eV, we get an increase in ${\rm Im} \epsilon$
\be
 \Delta({\rm Im}[\epsilon])= (\frac{9}{2.5})^2 
  \frac{1}{4.6\times10^{15} s^{-1} \times \  2.1\times 10^{-14}s}
  \approx 0.13.
\ee

\begin{figure}
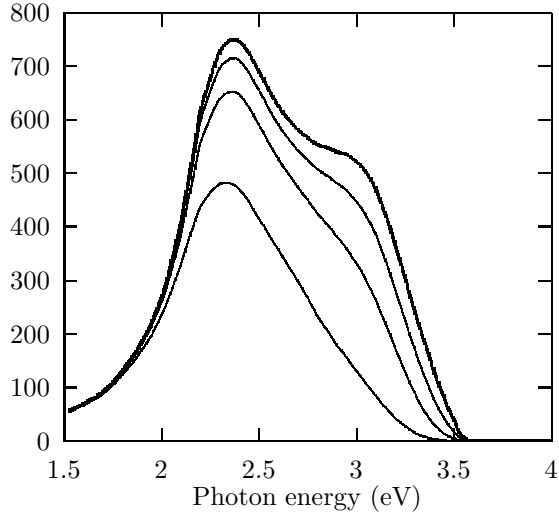

\caption{
Differential power 
(calculated with a white spectrum\protect\cite{white} 
for the driving current) for a Ag sample and 
Ir tip with $R$=300 \AA\  and $d$=5 \AA\@.
The different curves show how the spectrum develops as the 
imaginary part of the Ag dielectric function is increased:
The thick curve gives the result obtained with $\epsilon_{\rm Ag}$
determined from optical experiments. 
The remaining curves were calculated using
$\epsilon_{\rm Ag}+i0.1$,
$\epsilon_{\rm Ag}+i0.3$, and 
$\epsilon_{\rm Ag}+i1.0$,
respectively.
}
\label{Taufig}
\end{figure}

In Fig.\ \ref{Taufig} we show how an extra contribution (here taken to
be frequency-independent) to
${\rm Im} \epsilon$ affects the radiation spectrum for 
a Ag-sample-Ir-tip combination.
Extra damping leads to a decrease in the 
differential power, but the changes are still relatively small
for realistic modifications 
(i.e., $\epsilon_{\rm Ag}$+$i$0.1 and $\epsilon_{\rm Ag}$+$i$0.3)
of the dielectric function.
Maybe the best justification for using 
optically measured dielectric functions
is provided by a comparison between Figs.\ \ref{AgIrfig} and \ref{Taufig};
changes in the tip geometry have larger effects than surface scattering.

Next we estimate the damping rate that result from
the fact that a plasmon, which is a coherent superposition of 
electron-hole pair excitations, under certain circumstances can decay
into real electron-hole pairs.
To discuss this in a simple way, 
we will use the so called d-parameter theory of 
electromagnetic surface response of a 
jellium.\cite{Feib82,Apell,Liebsch,Plum}
The frequency-dependent (and complex) function $d_{\perp}(\omega)$
tells where the centroid of the induced screening charge 
is situated relative to the jellium edge. 
The order of magnitude of $d_{\perp}$ is set by $k_F^{-1}$,
i.e. $d_{\perp}\sim 1 $ \AA.
The surface response function of the jellium, 
$g(q_{\|},\omega)$,\cite{Surfresp}
is modified (to lowest order in $q_{\|}d_{\perp}$)
from its classical value $(\epsilon-1)/(\epsilon+1)$ 
to\cite{Feib82}
\be
 g(q_{\|},\omega)=\frac{\epsilon(\omega)-1}{\epsilon(\omega+1}
 \left[1+\frac{\epsilon(\omega)}{\epsilon(\omega)+1}\ 
  2q_{\|}d_{\perp}(\omega) \right].
\label{Surfresp}
\ee
Then, provided that $q_{\|}$ is large enough that retardation effects 
can be neglected ($q_{\|}\gg \omega/c$)
and using $\epsilon(\omega)=1-\omega_p^2/\omega^2$,
the surface plasmon at an isolated jellium surface has
the frequency 
\be
 \omega=\omega_s \left[1-q_{\|}d_{\perp}
 (\omega_s)/2 \right],
\label{suplasmcorr}
\ee
where the classical surface plasma frequency 
$\omega_s=\omega_p/\sqrt{2}$.
Thus, for large enough $q_{\|}$, the resonance frequency will 
have a shift away from $\omega_s$ depending on ${\rm Re}\ d_{\perp}$,
and the damping of the surface plasmon depends on ${\rm Im}\ d_{\perp}$.
The analysis leading to Eq.\ (\ref{suplasmcorr}) can be generalized
to the situation where two jellium surfaces, a distance $L$ apart,
are facing each other. 
The resonance frequency for the low-frequency interface plasmon (from which
the tip-induced mode illustrated in Fig.\ \ref{Bild} can be derived)
is given by
\be
 \omega_{\rm IP}=\omega_s\sqrt{1-e^{-q_{\|}L}}
 \left[1-\frac{1+e^{-q_{\|}L}}{1-e^{-q_{\|}L}}
  \frac{q_{\|}d_{\perp}(\omega_{\rm IP})}{2} \right],
\label{iplasmcorr}
\ee
if the two jellia are identical.

Let us apply this formula to the case of the tip-induced plasmon.
Using the parameter values
$\hbar\omega_{\rm IP}=2.5$ eV, $q_{\|}=\pi/100$ \AA$^{-1}$
corresponding to a typical plasmon diameter (cf.\ Sec.\ \ref{Solu}), 
and $L=5$ \AA,
we find that the electron-hole pair excitations will give an additional
full-width-half-maximum broadening 
$$ {\rm FWHM}_{\rm e-h\ pairs}=-2 {\rm Im}[\hbar\omega_{\rm IP}]\approx 
0.15\ {\rm eV}. $$
To get this value, we used 
$${\rm Im}\ d_{\perp}(\omega/\omega_p=0.3)
\approx 0.16 {\rm \AA},$$ 
as calculated by Liebsch\cite{Liebsch}
for $r_s=3$  appropriate for silver.
It is clear that the estimated FWHM is considerably smaller than the 
resonance peak widths found in Figs.\ \ref{AgIrfig} and \ref{Taufig}.

\section{Conclusions and outlook}
\label{Concsec}

In this paper, we have presented a  fully retarded 
calculation of the light emission
rate from a scanning tunneling microscope 
probing a noble metal surface.
The major aim was to investigate in what way, and to what degree,
the inclusion of retardation changed the results compared with
our earlier non-retarded calculations.\cite{JMA,ZPB}

In general, the results of the previous calculations 
still remain valid from a qualitative and semi-quantitative point of view.
Light emission is resonantly enhanced 
due to the formation of a tip-induced plasmon
mode in the cavity formed between the tip and sample. 
The enhancement amounts to 4 to 5 orders of magnitude compared with 
light emission in vacuum, and the quantum efficiency is 
typically 10$^{-4}$ photons per tunneling electron.
The  resonant photon energy is $\approx$2.1 eV for Au  samples
and $\approx$2.0 eV for Cu samples, i.e.,
essentially the same results as were found before.
In the case of Ag samples, taking retardation into account 
lowers the resonant photon energy by a few tenths of eV\@. 

Our results can also be compared with some earlier calculations
addressing either SERS,\cite{Take} or light emission.\cite{Madra}
Note that this comparison is possible thanks to the
reciprocity theorem.\cite{Reci}
The present results give spectra that are very similar to the ones
obtained in Ref.\ \onlinecite{Take}.
Also in terms of intensities, the
two calculations give similar results.
A direct comparison of the numbers is not possible;
here the relevant quantity is the square of the  field enhancement
at one point in space, while in SERS one calculates the fourth power
of the field enhancement averaged over a certain surface (a model sphere in
Ref.\ \onlinecite{Take}).

The results of a previous retarded calculation
by Madrazo, Nieto-Vesperinas, and Garcia\cite{Madra} 
addressing light emission from an STM
differ from ours on a couple of points:
(i) MNVG find lower enhancement factors, 10$^3$ at most, while we get
10$^4$ (Au sample--W tip), 10$^5$ (Ag--Ir), or even 10$^6$ (Ag--Ag).
With a Ag sample and a Ag tip, and {\em exactly the same} parameter
values for photon frequency, observation angle, and tip radius, 
and with $d=$8 \AA, we get an enhancement that is about 50 times larger
than that found by MNVG.
Moreover, MNVG state that a major part of the enhancement
(approximately a factor 100) originates from the  artificial corrugation 
of the sample.
An important reason for the lower enhancement appears to be their 
using a  two-dimensional geometry;
their model tip is a horizontal cylinder and not a sphere,
so the concentration of the electric field to the ``end'' of
the tip is much less accentuated in their calculation.
(ii) MNVG also found that interchanging tip and sample materials
changed the photon yield by as much as a factor of 20.  
They only found a considerable
enhancement when the sample material is a noble metal.
In the present calculation, interchanging tip and sample materials
changed the photon yield by at most 
a factor 2--3 (cf.\ Fig.\ \ref{Swapfig}).
Again, this shows that the electromagnetic tip-sample interaction only 
plays a minor role for the field enhancement within the model of MNVG,
whereas it is paramount in our model.

The present theory can be extended in a number of ways.
If the sample is covered by an overlayer, or the tip is covered by 
a spherical shell of a different material, the calculations can be 
carried through 
in essentially the same way after a few modifications have been made.
With some more effort it should also be possible to treat cases where
the tip has a more complicated shape, like for instance, a spheroid.
In principle also regularly corrugated or
randomly rough sample surfaces could be treated within an extension
of the present framework, but the amount of numerical computations 
would be considerably larger than now.
Finally, a calculation that could really treat an extended tip
(i.e.\ one that continues out towards a tip-holder) would be useful.
This is, however, a more difficult problem than those mentioned before.

In addition to this, the experiments by Alvarado and co-workers on
the polarization properties of the emitted light and its dependence 
on the surface and tip properties have raised some interesting questions.
For example, when the direction of magnetization of a Co film is reversed,
the degree of polarization of the emitted light 
can change by up to 20 \%.\cite{Alv}
For a magneto-optic effect, this is a very large number,
and the effect could be important if it turns out that it can
be used to study surface magnetism with good spatial resolution.
(However, in a similar experiment on iron, Pierce {\it et al}.\cite{NIST}
found essentially no light polarization due to magnetism.)
The experimental results  of Ref.\ \onlinecite{Alv}
indicate that tip-sample interactions play 
a role also for this effect.
The emitted light was more polarized in the tunneling regime 
(low bias) than in the field-emission regime.
Moreover, it is possible that the shape of the tip is important 
in these experiments.
A very high degree of polarization can be reached using an 
asymmetrically shaped tip on a non-magnetic surface.\cite{Santos2}

In summary, we have shown that retardation does play a role
in light emission from STM's, but that the changes are not too big.
Thus, theoretical studies of new, more complicated effects like the
light emission from magnetic surfaces can very well be carried 
out neglecting retardation.
Comparing our results with those of Ref.\ \onlinecite{Madra},
we conclude that it is essential to retain a three-dimensional
tip geometry in order to get an enhancement of the light emission rate 
that is comparable to those found in experiment.

\acknowledgments

This research was supported by the 
Swedish Natural Science Research Council (NFR)\@.
I thank Peter Apell and Richard Berndt for useful discussions and comments.

\appendix

\newpage

\section*{}

In this appendix we give explicit expressions for the functions
$f_l(\kappa)$, $g_l(\kappa)$,  $S(\kappa)$, and $s_l$;
we also indicate how these expressions are derived.

To this end it is useful to express the Green's function of 
Eq.\ (\ref{Greendef}) in 
representations suitable for cylindrical and spherical coordinates.
In cylindrical coordinates
\barr
 G(\rvec,\rpvec)&&=i\sum_{m=0}^{\infty}\frac{2-\delta_{m,0}}{4\pi}
 \cos{(m(\phi-\phi'))}
\breakeq
 \times
 \int_0^{\infty} \kappa d\kappa
 \frac{J_m(\kappa\rho)J_m(\kappa\rho')}{\sqrt{k^2-\kappa^2}}
 e^{i\sqrt{k^2-\kappa^2}|z-z'|},
\label{Gcyl}
\earr
and for spherical coordinates
\barr
 &&
 G(\rvec,\rpvec)
 =ik\sum_{l=0}^{\infty}
 j_l(kr^<)h_l(kr^>)
 \sum_{m=0}^{l}
 \frac{(2-\delta_{m,0})(2l+1)}{4\pi}
\breakeq
 \times
 \frac{(l-m)!}{(l+m)!}
 P_l^m(\cos{\theta})P_l^m(\cos{\theta'})
 \cos{(m(\phi-\phi'))},
\label{Gsph}
\earr
where $r^<(r^>)$ is the lesser (greater) of $r$ and $r'$, and $h_l$
is a spherical Hankel function.
In addition, the integral 
\barr
\int_0^{\pi}&&\sin{\theta}d\theta \ e^{i\sqrt{k^2-\kappa^2}R\cos{\theta}}
 J_1(\kappa R\sin{\theta})P_l^1(\cos{\theta})
 =
\breakeq
=2i^{l-1}P_l^1(\sqrt{1-\kappa^2/k^2})j_l(kR)
\label{Integ}
\earr
will be used repeatedly.

Let us begin by looking at the contributions to $\Bvec$ in Eq.\
(\ref{Kirchoff}) coming from the sample.
From Eq.\ (\ref{Bsamp}),
\be 
\Bvec(\rpvec)=\hat{\phi}'\int_0^{\infty}\kappa d\kappa
 B(\kappa)J_1(\kappa\rho')
\label{Bsurf}
\ee
at the sample surface.
To calculate the electric field just outside the sample surface,
we first
evaluate the $\Evec$ field 
{\em inside} the sample from
\be
 \Evec=\frac{ic}{\epsilon_1k} \nabla\times\Bvec,
\ee
and 
then use the fact that the tangential component of $\Evec$ is continuous 
across the sample surface. 
This yields
\be
 -\frac{ik}{c}(\hat{n}'\times\Evec(\rpvec))
 =
 \frac{i}{\epsilon_1}\hat{\phi}'\int_0^{\infty}
 \kappa d \kappa B(\kappa) J_1(\kappa\rho')\sqrt{k_1^2-\kappa^2}
\label{Esurf}
\ee
at the sample surface.
Inserting Eqs.\ (\ref{Bsurf}), (\ref{Esurf}), and (\ref{Gcyl})
into (\ref{Kirchoff}) and carrying out the surface integral,
we find that the sample surface contribution to 
$\Bvec(\rvec)$ everywhere in the vacuum region
can be written
\barr
 \Bvec^{\rm (s)}(\rvec)=
 \hat{\phi}\int_0^{\infty} \kappa d \kappa&&
 B(\kappa)
 J_1(\kappa \rho) 
 e^{i\sqrt{k^2-\kappa^2}(z-z_0)}
\breakeq
 \times
 \frac{1}{2}\left[1-
 \frac{\sqrt{k_1^2-\kappa^2}}{\epsilon_1\sqrt{k^2-\kappa^2}}
 \right]. 
\label{Bs}
\earr
Comparing this with Eq.\ (\ref{Bkeq}),
it is immediately clear that 
\be
 S(\kappa)=\frac{1}{2}\left[1-
 \frac{\sqrt{k_1^2-\kappa^2}}{\epsilon_1\sqrt{k^2-\kappa^2}} \right] .
\label{Skappaeq}
\ee
Furthermore, by calculating the overlap at the spherical surface 
between $\Bvec^{\rm (s)}(\rvec)$,
as given by Eq.\ (\ref{Bs}),
and $P_l^1(\cos{\theta})$, 
one finds, with the aid of Eq.\ (\ref{Integ}), the plane-to-sphere coupling
\barr
 f_l(\kappa)=&&\frac{2l+1}{l(l+1)}i^{l-1}
 S(\kappa)
\breakeq \times
 e^{i\sqrt{k^2-\kappa^2}(d+R)}
 P_l^1\left(\sqrt{1-\kappa^2/k^2}\right)j_l(kR).
\earr
Note that if $\epsilon_1=1$ both $S(\kappa)$ and $f_l(\kappa)$ vanish.

Next, we apply the same methods to calculate the contributions
to $\Bvec(\rvec)$ from the spherical surface.
This yields
\barr
 s_l&&=ikR\ h_l(kR)\ j_l(kR)
\breakeq
\times
 \left(
 kR\ \frac{j_l'(kR)}{j_l(kR)}+1
 -\frac{1}{\epsilon_2}\ \left[k_2 R\ \frac{j_l'(k_2 R)}{j_l(k_2 R)}+1 
 \right]
 \right) 
\label{sleq}
\earr
for the sphere-sphere coupling, while
\barr
 &&g_l(\kappa)=\frac{i^l\ R^2}{\sqrt{k^2-\kappa^2}}
 \ e^{i\sqrt{k^2-\kappa^2}(d+R)} \ 
 P_l^1\left(\sqrt{1-\kappa^2/k^2}\right)
\breakeq
\times
 \left(
 k\frac{j_l'(kR)}{j_l(kR)}+\frac{1}{R}-
  \frac{1}{\epsilon_2}
 \left[\frac{1}{R}+k_2\frac{j_l'(k_2 R)}{j_l(k_2 R)}\right]
 \right)
  j_l(kR)
\earr
describes how the waves originating from the sphere propagates to 
the sample surface.
In this case, $\epsilon_2=1$ causes $s_l$ and $g_l(\kappa)$ to vanish.

\newpage
\end{multicols}


\medskip \medskip
\medskip
\input fig2.tex

\newpage

\medskip
\medskip
\input fig3.tex

\medskip
\medskip
\input fig4.tex

\medskip
\medskip
 \input fig5.tex

\medskip
\medskip
\input fig6.tex

\medskip
\medskip
\input fig7.tex

\medskip
\medskip
\input fig8.tex

\medskip
\medskip
\input fig9.tex

\medskip
\medskip
\input fig10.tex


\begin{references}
\bibitem{white} To be precise, this means that we have used 
 a source of the form in Eqs.\ (\ref{Modelcurr}) and (\ref{Form}), where
 $C=1$ A/\AA\ (and independent of the frequency) and $a=5$ \AA\@.   
 Thus, the different results of this paper
 obtained with a white-spectrum source can be compared with
 each other directly.
\bibitem{Agplasm} This value is set by the density of conduction electrons,
     and describes the optical properties of Ag for 
     $\hbar \omega \alt 2.5$ eV reasonably well.
   The experimental plasmon energy, 3.8 eV, is strongly 
   influenced by interband transitions.
\end{references}
\end{document}